\documentstyle[12pt]{article}
\title{
Fermions in the vortex core in chiral superconductors.}
\author{ G.E. Volovik\\
Low Temperature Laboratory\\
Helsinki University of Technology\\
P.O.Box 2200, FIN-02015 HUT, Finland\\
and\\
L.D. Landau Institute for Theoretical Physics, \\
Kosygin Str. 2, 117940 Moscow, Russia\\
}
\begin{document}
\maketitle
\begin{abstract}
{Using the semiclassical approach to the energy levels of the fermions bound
to the vortex core, we found the difference between the states in nonchiral
and chiral superconductors, determined by the Berry phase. The bound states of
fermions in the singly quantized vortex in the layered superconductor with the
symmetry of superfluid $^3$He-A is $E=n\omega_0$ (as distinct from
$E=(n+1/2)\omega_0$ in $s$-wave superconductor) and thus contains the state
with exactly zero energy. This is in accordance with the result obtained in
microscopic theory \cite{KopninSalomaa}.  Using this approach for calculations
of the effect of impurities on the spectrum of bound state we reproduced the
Larkin-Ovchinnikov result for single impurity in $s$-wave vortex
\cite{Larkin-Ovchinnikov,KoulakovLarkin2}: the spectrum  has the double period
$\Delta E=2\omega_0$ and consists of two equdistant sets of levels. The same
approach however shows that the single impurity does not
change the spectrum $E=n\omega_0$ in the $p$-wave vortex .
 }
\end{abstract}

\section{Introduction}

The low-energy fermions bound to the vortex core play the main role in the
thermodynamics and dynamics of the vortex state in superconductors and
Fermi-superfluids. The spectrum of the low-energy bound states in the core
of the
axisymmetric vortex with winding number $N=\pm 1$ in the isotropic model of
$s$-wave superconductor was obtained by Caroli, de Gennes and Matricon
\cite{Caroli}:
\begin{equation}
E_n=-N\omega_0\left(n+{1\over 2}\right)~,
 \label{Caroli}
\end{equation}
This spectrum is two-fold degenerate due to spin degrees of freedom. The
integral quantum number $n=L_z$ is a modified  angular momentum of the
bound state fermions, while the level spacing $\omega_0$ is the magnitude
of the
angular velocity of the fermions orbiting about the vortex axis. The
direction of
rotation is determined by the sign of the winding number $N$ of the vortex.

The level spacing is small compared to the energy gap of the quasiparticles
outside the core, $\omega_0\ll\Delta$. So, in many physical cases the
descreteness of $n$ can be neglected. In such cases the spectrum crosses zero
energy as a function of continuous angular momentum $L_z$. So, one has the
fermion zero modes. The fermions in this 1D "Fermi liquid" are chiral: the
positive energy fermions have a definite sign of the angular momentum $n$.

The motion of the vortex perturbs the fermions in the vortex core and leads to
the spectral flow  of the fermionic levels from the negative energy vacuum
to the
positive energy world of excitations forming the heat bath, or the normal
component. This spectral flow leads to the momentum exchange between the vortex
core and the heat bath and thus to the additional force acting on the vortex.
when it moves with respect to the heat bath
\cite{Q-modes-Index,KopninVolovikParts,Stone,Bevan}. This force was calculated
in the microscopic theory in the earlier papers by Kopnin and coathors
\cite{KopninCoAuthors}.

Later the other types of vortices have been found to exist in superfluid $^3$He
and in high-temperature superconductors: with different winding numbers and in
the nonsymmetric environment and/or with the spontaneously broken symmetry in
the core. The microscopic description of the bound-state fermions in such
distorted vortices becomes complicated. So one needs in the simple
phenomenological theory, which describes the low-energy motion of the
quasiparticles in the vortex core in the same manner as the low-energy fermions
are described in the Landau theory of the Fermi liquid.

Such theory was constructed in the papers
\cite{Stone,KopninVolovik,RotatingCore}. In this approach the fast radial
motion of the fermions in the asymmetric vortex core is integrated out and one
obtains only the slow motion corresponding to the fermions zero modes. The slow
low-energy dynamics is described quasiclassically in terms of the pair of
canonically conjugated variables: the angle $\theta$, which specifies the
direction of the linear momentum of the propagating fermion, and impact
parameter (or angular momentum
$L_z$).  In this description  the fermion zero modes are represented by
one or several bracnches $E_{\rm a}(L_z,\theta)=-\omega_{\rm a}(\theta)
(L_z-L_{z{\rm a}}(\theta))$ of the fermionic spectrum. They cross zero
energy as
a function of the continuous angular momentum $L_z$. The number of fermion zero
modes does not depend on $\theta$ and is determined by the vortex winding
number
$N$
\cite{Q-modes-Index}: it is $2N$ if the spin degrees of fredom are taken into
account. The system of these chiral fermions form the set of the one
dimensional
Fermi-liquids.

The final step in the calculation of the quantum spectrum of the
fermions in the asymmetric core is the quantization of the remaining slow
motion.
It is obtained by the Bohr-Sommerfeld rule for the canonically conjugated
variables
\cite{KopninVolovik,RotatingCore}. It appears that even for the asymmetric
vortex the main structure of the quantum energy levels remains robust. One
obtains  one or several branches of the discrete energy levels. On each branch
the levels are equidistant as in the simplest case of the axisymmetric
vortex in
$s$-wave superconductor.

Here we show that the Berry phase is important for quantization. Due to the
Berry phase  the bound states of
fermions in the singly quantized vortex in the chiral layered superconductor
with the symmetry of superfluid $^3$He-A (possible candidate is
 superconducting Sr$_2$RuO$_4$ \cite{Rice}) is $E=n\omega_0$, while the
$s$-wave superconductor one has
$E=(n+1/2)\omega_0$. Thus the spectrum of the $p$-wave vortex contains the
state
with exactly zero energy, which confirms the
microscopic calculations in \cite{KopninSalomaa}. We also consider the
effect of
impurities on the spectrum of bound state. We reproduce the Larkin-Ovchinnikov
result for single impurity in $s$-wave vortex
\cite{Larkin-Ovchinnikov,KoulakovLarkin2} and extend it to the $p$-wave case.

\section{Boundary conditions for quantization of bound states in the
core of $s$-wave vortex.}

Let us start with the $N$-quantum nonaxisymmetric vortex in conventional
superconductors, where the asymmetry is caused, say, by crystal potential or by
external perturbation.
In the semiclassical approximation the quasiparticle states are
characterized by their trajectories. Trajectories are straight
lines along the direction of the linear momentum in transverse plane   ${\bf
q}_\perp$, with the total momentum being on the Fermi surface,  $q_z^2+{\bf
q}_\perp^2=p_F^2$. Let us consider somewhat more general case of distorted
vortex in
$s$-wave superconductor. The order parameter is chosen as
$ \Delta(r,\phi)=|\Delta(r,\phi)|e^{i\Phi(\phi)}$.
Here $\phi$ is azimuthal angle;
$\Phi(\phi)$ is the phase of the order parameter, which winding around the
vortex is
\begin{equation}
N={1\over 2\pi}\int_0^{2\pi} d\phi {d\Phi\over d\phi}~,~\Phi(\phi)=N\phi +
\tilde\Phi(\phi)~,
 \label{Winding}
\end{equation}
where $\tilde\Phi(\phi)$ is a periodic function of $\phi$. The Bogoliubov
Hamiltonian for quasiparticles in the vocinity of such a vortex is
\begin{equation}
{\cal H}={\bf p}\cdot{\bf v} \tau_3 +
\tau_1|\Delta(r,\phi)|\cos \Phi(\phi) +
\tau_2|\Delta(r,\phi)|\sin \Phi(\phi)~,~{\bf v}={{\bf q}_\perp\over m}~.
 \label{MicroHamiltonian}
\end{equation}

Let us first neglect the conventional scattering on vortices, leaving only
the Andreev reflection. In this case, in the quasiclassical limit there is
no transition between different trajectories: quasiparticle is moving along the
same trajectory, i.e. with the same ${\bf
q}_\perp$,  and experience the Andreev scattering which does not change the
momentum, but chnges the sign of the group velocity. Each trajectory is
characterized by the angle
$\theta$ in the transverse plane and by the impact parameter $b$:
\begin{equation}
b=r\sin  \phi ~,~ {\hat{\bf q}}_\perp = \hat{\bf x} \cos  \theta + \hat{\bf y}
\sin \theta~.
 \label{AngleImpact}
\end{equation}

Let us make two transformations of the Eq.(\ref{MicroHamiltonian}) for given
${\bf q}_\perp$. The first one is the coordinate transformation which orients
the axis $x$ along the trajectory ${\bf q}_\perp$:
\begin{equation}
{\cal H}=  -iv\tau_3 \partial_s+
\tau_1|\Delta(r,\phi-\theta)|\cos \Phi(\phi-\theta) +
\tau_2|\Delta(r,\phi-\theta)|\sin \Phi(\phi-\theta)~.
 \label{MicroHamiltonian2}
\end{equation}
In a new coordinate system $p_x=-i\partial_s$, where $s=r\cos\phi$ is the
coordinate along the trajectory.

The second transformation is intended to make the phase of the order parameter
the periodic function of $\theta$:
\begin{eqnarray}
\Psi\rightarrow
e^{i N\tau_3 \theta/2}\Psi~,\label{TransformationFunction}\\
{\cal H}\rightarrow
e^{i N\tau_3 \theta/2}{\cal H}e^{-i N\tau_3 \theta/2}=\\
  -iv\tau_3 \partial_s +
\tau_1|\Delta(r,\tilde\phi)|\cos \left(\tilde\Phi(\tilde\phi) +N\phi\right)+
\tau_2|\Delta(r,\tilde\phi)|\sin \left(\tilde\Phi(\tilde\phi) +N\phi\right)~,
 \label{TransformationHamiltonian}
\\
\tilde\phi=\phi-\theta~.
\end{eqnarray}

Note that for the axisymmetric vortex there is no more dependence on $\theta$:
\begin{equation}
{\cal H}=
  -iv\tau_3 \partial_s +
\tau_1|\Delta(r)|\cos \left(N\phi\right)+
\tau_2|\Delta(r)|\sin \left(N\phi\right)~.
\label{MicroHamiltonianSymmetric}
\end{equation}
The
quantity
$N\theta/2$ plays the part of the Berry phase: when one changes the direction
of the trajectory, the only change in the wave function is multiplication by
$e^{i N\tau_3 \theta/2}$. In more general case of nonsymmetric vortex this
Berru phase is the adiabatic change in the phase of the wave function, when one
continuously rotates the trajectory of the quasiparticle by the angle
$\theta$ in the transverse plane.

Quantization of the motion along $s$ gives rise to generally $N$ fermion zero
modes of the type:
\begin{equation}
E(L_z,\theta)=-\omega_0(\theta)\left(L_z-L_z^{(0)}(\theta)\right)
 \label{QuasiclassicalEnergy}
\end{equation}
 The next step is quantization of the azimuthal motion. Here it is
important that the wave function in terms of $\theta$ has the boundary
condition
\begin{equation}
\Psi(\theta +2\pi)=(-1)^N\Psi(\theta)
 \label{BoundaryCondition}
\end{equation}
  which follows from
Eq.(\ref{TransformationFunction}).

The shift $L_z^{(0)}(\theta)$ from zero is
antisymmetric because of the "CPT"-theorem:
\begin{equation}
L_z^{(0)}(\theta)=-L_z^{(0)}(\theta +\pi)
 \label{ShiftCondition}
\end{equation}
Below Eqs(\ref{QuasiclassicalEnergy},\ref{ShiftCondition}) will be explicitly
shown for a weakly asymmetric vortex.

The quantum Hamiltonian for the canonically conjugated variables
$L_z=-i\partial_\theta$ and
$\theta$ is
\begin{equation}
{\cal H}= - {1\over 2} \left\{ ~\omega_0(\theta)~,~\left(-i{\partial
\over
\partial\theta} - L_z^{(0)}(\theta)\right)\right\}~~ ,
 \label{HamiltonianForAngle}
\end{equation}
where $\{~,~\}$ is anticommutator. The normalized wave function of the eigen
state with the energy $E$ is
\begin{equation}
\Psi(\theta)=\left<{1   \over
\omega_0(\theta)}\right>^{-1/2} ~ {1\over
\sqrt{2\pi \omega_0(\theta)}}\exp \left(i\int^\theta d\theta'\left( {E\over
\omega_0(\theta')} + L_z^{(0)}(\theta')\right) \right)~~.
 \label{Solution}
\end{equation}
The energy eigenvalues ${\cal H}\Psi_n(\theta)=E_n\Psi_n(\theta)$ are found
from
the requirement that according to Eq.(\ref{BoundaryCondition}) the wave
function
$\Psi_n(\theta)$ changes sign after encircling the origin in the momentum space
$\theta$ for odd $N$ and does not change if $N$ is even, i.e. the
phase of the wave function changes by
$\pi (2n+N)$. This gives
\begin{equation}
E_n  \int^{2\pi}_0 { d\theta\over
\omega_0(\theta)}+ \int^{2\pi}_0   d\theta L_z^{(0)}(\theta) = -2\pi \left(n
+{N\over 2}\right)  ~~.
\label{EigenValues}
\end{equation}
The second term in the lhs of Eq.(\ref{EigenValues}) is zero according to
Eq.(\ref{ShiftCondition}), and one obtains the quantization which is
essentially the same as for the axisymmetric vortex:
\begin{eqnarray}
E_n = - {n +{1\over 2}\over \left<
\omega_0^{-1}(\theta)\right> }
  ~~,~~{\rm odd}~~N\\
E_n = -  {n\over \left< \omega_0^{-1}(\theta)\right> }
   ~~,~~{\rm even}~~N.
\label{EigenValues2}
\end{eqnarray}
For even $N$ there is a state with zero energy.

\section{Fermion zero modes in a weakly asymmetric vortex with odd $N$ in
$s$-wave superconductors.}

Let us consider a weakly asymmetric vortex in which the anisotropy is
small, say,
$|\tilde\Phi(\tilde\phi)|\ll 1$.   In the extreme limit, when $b=0$
and $\tilde\Phi(\tilde\phi)=0$, the trajectory  with $\phi=0$ for $s>0$ and
$\phi=\pi $ for $s<0$ gives exactly zero energy if $N$ is odd. This is
becuase the order parameter along this trajectory is real and changes sign
when the origin is crossed. The index theorem requires that there is the
normalizable eigen state with zero energy.

Let us choose an odd $N$, then for small $b$ the perturbation theory can be
constructed. Since $\phi$ is close to $0$ or $\pi$ one has
\begin{equation}
\tilde\phi=\phi -\theta \approx {\pi\over 2} \left(1-{\rm sign}~ s\right)-
\theta~.
 \label{Approx}
\end{equation}
and
\begin{eqnarray}
{\cal H}=
 {\cal H} ^{(0)} +{\cal H} ^{(1)}~,\\
{\cal H} ^{(0)} =-iv \tau_3 { d\over ds} +
\tau_1|\Delta(s,\theta)| ~{\rm sign~ s}~,
\\
\Delta(s,\theta)=\Delta\left(|s|,{\pi\over 2} \left(1-{\rm sign}~
s\right)- \theta\right)
\\
{\cal H}^{(1)}=\tau_2|\Delta(s,\theta)|\sin \left(\tilde\Phi(\tilde\phi)
+N\phi\right)\approx  \\
\approx \tau_2|\Delta(s,\theta)| \left( N{b\over|s|} +
~({\rm sign~ }s)~ \tilde\Phi\left({\pi\over 2} \left(1-{\rm sign}~
s\right)- \theta\right)
 \right)~.
 \label{Perturbations}
\end{eqnarray}
The Hamiltonian ${\cal H} ^{(0)}$ has fermion zero mode, $\Psi^{(0)}(s)$,  so
that the energy is the average of ${\cal H} ^{(1)}$ over this mode:
\begin{eqnarray}
E(b,\theta)=\left<{\cal H}^{(1)}\right>_s=\nonumber\\
-N
b\left<{|\Delta(|s|,-\theta)| +|\Delta(|s|,\pi-\theta)|\over 2|s|}\right>_s +
\label{Energy1}\\ +\left< |\Delta(|s|,-\theta)|  \tilde \Phi(-\theta)
-|\Delta(|s|,\pi-\theta)|\tilde
\Phi(\pi -\theta)\right>_s ~.
 \label{Energy2}
\end{eqnarray}
This gives the Eq.(\ref{QuasiclassicalEnergy}) for the quasiclassical
energy $E(L_z,\theta)$ and the Eq.(\ref{ShiftCondition}) for
the shift $L_z^{(0)}(\theta)$. The Eq.(\ref{Energy2}) is
$E(0,\theta)=\omega_0(\theta)L_z^{(0)}(\theta)$: it is the energy of
quasiparticle moving along trajectory with
$b=0$. The nonzero
value of this energy comes
from the imaginary part of the order parameter ${\rm Im} ~\Delta(|s|,-\theta)$
on the trajectory.

\section{Boundary conditions for quantization of bound states in the vortex
core in chiral superconductor.}

Let us consider simplest case of the axisymmetric vortex in $p$-wave
superconductor with the $^3$He-A order parameter. The order parameter in this
vortex is
$c(r)(p_x+ip_y)e^{iN\phi}$, where $c={\Delta_0\over p_F}$ and $\Delta_0$ is
the gap amplitude. The Bogoliubov Hamiltonian for quasiparticles in the vortex
core along the trajectory with given $\theta$ is
\begin{equation}
{\cal H}={\bf p}\cdot{\bf v} \tau_3 +
 \tau_1  q_\perp c(r)   \cos(\theta + N\phi)  +\tau_2
q_\perp  c(r)
\sin(\theta  + N\phi) ~,~c(r)={\Delta_0(r)\over p_F}~.
 \label{MicroHamiltonianP}
\end{equation}

The coordinate transformation which orients
the axis $x$ along the trajectory ${\bf q}_\perp$:
\begin{equation}
{{\cal H}\over   q_\perp}= -i{1\over m}\tau_3 \partial_s +
 \tau_1  c(r)   \cos \left(N \phi -(N-1)\theta)\right)
+\tau_2     c(r)
  \sin \left(N \phi -(N-1)\theta)\right)  ~.
 \label{MicroHamiltonian2P}
\end{equation}

The second transformation is intended to delete the dependence on  $\theta$:
\begin{eqnarray}
\Psi\rightarrow
e^{i (N-1)\tau_3 \theta/2}\Psi~,\label{TransformationFunctionP}\\
e^{i (N-1)\tau_3 \theta/2}{\cal H}e^{-i (N-1)\tau_3 \theta/2}=
 q_\perp\left( -{i\over m}\tau_3 \partial_s +\tau_1  c(r)   \cos  N \phi
+\tau_2     c(r) \sin  N \phi \right)  .
 \label{TransformationHamiltonianP}
\end{eqnarray}

The new Hamiltonian is the same as for the $N$-quantum symmetric vortex
in $s$-wave superconductor, but the Berry phase is different, now it is
$(N-1)\tau_3 \theta/2$. This leads to different boundary condition for
the wave function, which according to Eq.(\ref{TransformationFunctionP}) is
\begin{equation}
\Psi(\theta +2\pi)=(-1)^{N+1}\Psi(\theta)
 \label{BoundaryConditionP}
\end{equation}
This leads to the following quantization of levels in the core of the $p$-wave
vortex with odd $N$:
\begin{equation}
E_n = - n \omega_0
  ~~,~~{\rm odd}~~N~.
\label{EigenValuesP}
\end{equation}
Thus for $N=\pm 1$ vortex there is a zero energy level.

In general case  the Cooper pair has an angular momentum projection $m$.
Before we considered two examples: the $s$-wave superconductor, which
belongs to the class $m=0$; and the A-phase-like superconductors, which
belongs to class $m=\pm 1$. In the case of general $m$, the
Berry phase and boundary condition for the wave function of fermions in
$N$-quantum vortex are
\begin{equation}
\Theta_{\rm Berry}={N-m\over 2}\tau_3 \theta ~~,~~\Psi(\theta
+2\pi)=(-1)^{N-m}\Psi(\theta)
 \label{BoundaryConditionGen}
\end{equation}
This leads to two classes of the fermionic spectrum in the symmetric core:
$E_n=n\omega_0$ if $N-m$ is even and $E_n=(n+1/2)\omega_0$ if $N-m$ is odd.

\section{Effect of impurity scattering.}

Elastic scattering on impurity causes the transition between different
trajectories. In the limit of low energy of the quasiparticle the impact
parameter of of the scattered particle tends to zero and becomes smaller than
the distance from impurity to the center of the vortex. If the size of
impurity is small then the scattering of the low-energy quasiparticle occurs
only between the trajectories along the line between the vortex and
impurities, i.e. between $\theta=\theta_{\rm imp}$ and   $\theta=\pi
- \theta_{\rm imp}$ \cite{KoulakovLarkin1}. The "Josephson coupling" between
these two states is
\begin{equation}
{\cal H}_{\rm imp}=2\lambda e^{i\gamma}\Psi(\pi- \theta_{\rm
imp})\Psi^*(\theta_{\rm imp}) + 2\lambda e^{-i\gamma}
\Psi^*(\pi-
\theta_{\rm imp})\Psi(\theta_{\rm imp})~.
 \label{JosephsonCoupling}
\end{equation}

The Schr\"odinger equation is now
\begin{equation}
-i\omega_0{\partial \Psi
\over
\partial\theta} +2\lambda e^{i\gamma} \delta(\theta-\theta_{\rm imp})\Psi(\pi-
\theta_{\rm imp}) +2\lambda e^{-i\gamma} \delta(\theta-\pi+\theta_{\rm
imp})\Psi(\theta_{\rm imp}) =E\Psi(\theta)~~ ,
 \label{HamiltonianImpurity}
\end{equation}

Let us choose the position of impurity at $\pi/2$. Then the relevant
trajectories, where the scattering occurs, are at $\pm \pi/2$ and one has
\begin{equation}
-i\omega_0{\partial \Psi
\over
\partial\theta} +2\lambda e^{i\gamma} \delta\left(\theta-{\pi\over
2}\right)\Psi\left(-{\pi\over 2}\right) +2\lambda e^{-i\gamma}
\delta\left(\theta+{\pi\over 2}\right)\Psi\left({\pi\over 2}\right)
=E\Psi(\theta)~~ ,
 \label{HamiltonianImpurity2}
\end{equation}
with boundary condition
\begin{equation}
 \Psi\left(-\pi\right) =\pm \Psi\left( \pi \right)
 \label{Boundary2}
\end{equation}
Here the sign $+$ is for the $p$-wave $N=1$ vortex and $-$ is for $s$-wave
$N=1$
vortex.
The solution is
\begin{eqnarray}
\Psi =A_1 e^{i{E\over \omega_0}\theta} ~~,~~-{\pi\over 2}<\theta<{\pi\over
2}~,\\
\Psi =A_2 e^{i{E\over \omega_0}\theta} ~~,~~-\pi<\theta<-{\pi\over 2}~,
\\
\Psi =A_3 e^{i{E\over \omega_0}\theta} ~~,~~{\pi\over 2}<\theta<\pi
~.
 \label{Solution}
\end{eqnarray}

The conditions across two $\delta$-function potentials and the boundary
condition Eq.(\ref{Boundary2}) give 3 equations for the parameters $A$:
\begin{eqnarray}
-i\omega_0 e^{-i{\pi E\over 2\omega_0} } (A_1 - A_2)= \lambda  e^{i
\gamma} e^{i{\pi E\over 2 \omega_0} }  (A_1+A_3)~~,
 \\-i\omega_0 e^{i{\pi E\over
2\omega_0} } (A_3 - A_1)= \lambda  e^{-i \gamma}e^{-i{\pi E\over 2 \omega_0} }
(A_1+A_2)~,
\\
  e^{ i{\pi E\over  \omega_0} }A_3=\pm  e^{ -i{\pi E\over  \omega_0} }A_2~.
 \label{A}
\end{eqnarray}
Solution of these equations give the energy eigenvalues:
\begin{eqnarray}
\cos {\pi E\over  \omega_0} = {2\omega_0 \lambda\over \omega_0^2+
\lambda^2}\sin\gamma ~~,~~s-{\rm wave}~, \label{EigenValuesS}
 \\
\sin {\pi E\over  \omega_0} = {2\omega_0 \lambda\over \omega_0^2+
\lambda^2}\cos\gamma ~~,~~p-{\rm wave}~.
 \label{EigenValuesP}
\end{eqnarray}
In the $s$-wave case the Eq.(\ref{EigenValuesS}) is similar to Eq.(2.10) of
Ref.\cite{KoulakovLarkin2}: the spectrum in the presence of impurity has the
double period $\Delta E=2\omega_0$ and consists of two equdistant sets of
levels. These two sets transform to each other under symmetry transformation
$E\rightarrow -E$, which is the "CPT"-symmetry of the system.

For the $p$-wave case the Eq.(\ref{EigenValuesP}) also gives two sets of
levels with the alternating shift. But the two sets are not mutually symmetric
with respect to $E=0$. This contradicts to  the "CPT"-symmetry of the
system. The only way to reconcile the Eq.(\ref{EigenValuesP}) with the symmetry
is to fix the phase $\gamma=\pi/2$ of the  "Josephson coupling". Then the
energy levels are $E_n=n\omega_0$, i.e.  the same as without impurities. Thus
the same "CPT"-symmetry, which is responsible for the eigenstate with $E=0$,
provides the rigidity of the spectrum.

\section{Conclusion}

The Berry phase in Eq.(\ref{BoundaryConditionGen}) is instrumental for the
Bohr-Sommerfeld quantization of the energy levels in the vortex core. It
chooses between the two possible quantizations consistent with the
$CPT$-symmetry of states in superconductors:   $E_n=n\omega_0$ and
$E_n=(n+1/2)\omega_0$.

We found that the two spectra remain intact if the small perturbation is
added, which violates the axial symmetry of the order parameter. We also
considered the effect of single impurity on the spectrum of bound states. We
found that if in a pure superconductor the spectrum is
$E_n=n\omega_0$ (an example is $N=1$ vortex in chiral superconductor with
$m=1$) then the impurity does not change this spectrum. On the other hand if
the initial spectrum is
$E_n=(n+1/2)\omega_0$, the impurity splits it into two series according to the
Larkin-Ovchinnikov prescription
\cite{Larkin-Ovchinnikov,KoulakovLarkin2}.
This rigidity of the spectrum must be taken into account when the effect of
randomness due to many impurities is considered with introduction of new
level statistics for the fermionic spectrum in the core
\cite{FeigelmanSkvortsov}.

I am indebted to P. Wiegmann who noticed that there should be a Berry phase
whcih is reponsible for quantization and to M. Feigel'man for fruitful
discussions.

\end{document}